\documentclass[preprint,12pt]{elsarticle}

\usepackage{amssymb}
\usepackage{graphicx}
\usepackage{subfigure}
\usepackage{multirow}
\usepackage{amsmath}
\usepackage{geometry}
\usepackage{tabularx}
\usepackage{amsmath}          
\usepackage{amssymb}
\usepackage{bm}
\usepackage{cite}
\usepackage[title]{appendix}
\usepackage{xcolor}
\usepackage{booktabs}
\journal{Elsevier}
\begin{document}

\begin{frontmatter}



\title{Stochastic additive manufacturing simulation: from experiment to surface roughness and porosity prediction}


\author[inst1]{Yangfan Li}

\affiliation[inst1]{organization={Department of Mechanical Engineering},%Department and Organization
            addressline={Northwestern
University}, 
            city={Evanston},
            postcode={60208}, 
            state={IL},
            country={USA}}

\author[inst1]{Ye Lu\corref{cor1}}
\ead{ye.lu@northwestern.edu}
\author[inst1]{Abdullah Al Amin}
\author[inst1]{Wing Kam Liu\corref{cor1}}
\ead{w-liu@northwestern.edu}
\cortext[cor1]{Co-corresponding authors}

\begin{abstract}
Deterministic computational modeling of laser powder bed fusion (LPBF) process fails to capture irregularities and roughness of the scan track, unless expensive powder-scale analysis is used. In this work we developed a stochastic computational modeling framework based on Markov Chain Monte Carlo (MCMC) capable of capturing the irregularities of LPBF scan. The model is calibrated against AFRL single track scan data using a specially designed tensor decomposition method, i.e., Higher-Order Proper Generalized Decomposition (HOPGD) that relies on non-intrusive data learning and construction of reduced order surrogate models.Once calibrated, the stochastic  model can be used to predict the roughness and porosity at part scale at a significantly reduced computational cost compared to detailed powder-scale deterministic simulations. The stochastic simulation predictions are validated against AFRL multi-layer and multitrack experiments and reported as more accurate when compared with regular deterministic simulation results. 

\end{abstract}




\end{frontmatter}


\section{Introduction}
\label{sec:sample1}

Even though Laser powder bed fusion (L-PBF), a promising Additive Manufacturing (AM) process is widely used in the aerospace, automotive, and biomedical industries \citep{doi:10.1080/00207540903479786,Yan2012EvaluationsOC,guo2013additive}, the processing-structure-properties-performance (PSPP) relationship
must be understood before the industry can rely upon AM to produce components with consistent mechanical properties for parts under fatigue loading. Fatigue failure, one of the most common damage modes in cyclically loaded metallic materials, is influenced by defects within components inherently created by laser powder bed fusion (L-PBF) Additive Manufacturing (AM) processes. AM inherently created defects such as surface roughness and porosity, which may occur due to improper melt pool formation from insufficient melting caused by too little energy absorption or trapped gas caused by vaporization \citep{cunningham2017synchrotron,tang2017prediction}, are driving factors in the fatigue performance of AM components \citep{yadollahi2017additive}.To choose the optimal manufacturing workflow, one must understand the surface roughness capabilities of metal AM, as well as post-processing techniques and their associated time and cost. The reduction of porosity also reduces the risk of cracks and fatigue fractures. The surface roughness and porosity of a part is critical to its function and long-term performance \citep{whip2019effect}. However, the prediction of surface roughness and porosity generally requires detailed simulations and is time-consuming \citep{goh2021review} and therefore rarely used for large component scale analysis. Thus, more efficient approaches should be introduced into AM simulation models to find the relations between computation inputs and the mechanical performance of AM-built parts. 

The formation of a melt pool, which is an intermediate step between solidification and laser source absorption, plays an important role in the description of interactions between powder materials \citep{witherell2014toward}.  In L-PBF, the localized solid powder is heated up and melted into a liquid after absorbing energy from the passing laser, then cools down and solidifies into a bulk material with a resultant microstructure as the laser moves further away. The melt pool acts as an effective index for the overall product quality since its geometrical characteristics, as one of the most important process signatures developed during an AM process \citep{yeung2020meltpool}, have huge influence on various properties of AM parts. A good predictive capability for the melt pool geometry can help on an efficient control of surface roughness. Besides, the melt pool cross-sectional area, which is typically measured by the pool width and depth, largely determines the porosity formation due to insufficient pool overlapping \citep{qiu2019influence}. Thus, the computational model that manipulates process parameters such as laser power, scan speed, and spot size can be built to control the melt pool as well as the corresponding mechanical behaviors \citep{11959300075f47619b6540a9764ebd85}.

Computational models have been widely used in additive manufacturing due to their high flexibility and efficiency, and many efforts on understanding the influence of process parameters on part quality are made. Finite element/volume method-based thermal model, which defines as a highly transient manufacturing process, is one of the most popular means. For example, a finite-volume-based simulation model is built by Ghosh et al. and is validated through experimental melt pool geometries with multiple laser power and scan speed combinations \citep{ghosh2018single}. A thermodynamically consistent model is also proposed for microstructure evolution during AM process using finite element method \citep{smith2016thermodynamically}. However, the lack of consideration on fluid flow in the melt pool neglects the effects of cooling through fluid convection, and makes the above models less accurate than thermal-fluid flow models when predicting the melt pool geometries \citep{yan2020data}. Among different thermal fluid models, Gan et al. proposed a well-tested transient three-dimensional thermal-fluid computational model to predict the thermal field in the entire part and velocity field in the melt pool region \citep{11959300075f47619b6540a9764ebd85}. The model is calibrated through highly controlled experiments made by Additive Manufacturing (AM) Modeling Challenge Series in 2020 \citep{schwalbach2021afrl} to validate and ensure the accuracy. Although the above model has accurately predicted melt pool geometries, the lack of stochastic information limits its capacity for predicting parts surface defects (e.g., surface roughness \citep{zhao2018effect})  and volumetric structural defects (e.g., porosity \citep{zhang2021microstructure}). 

The defects in AM product quality are mostly caused by uncertainty from various uncertainty sources existing in the complex AM process \citep{hu2017uncertainty}. Uncertainty quantification (UQ) has attracted tremendous interest in many research areas including AM in order to dramatically improve product reliability and understand the intrinsic uncertainty associated with the computational model \citep{hu2017uncertainty,li2019improved}. Typical uncertainty sources include natural variation in powder absorptivity, fluctuation in temperature boundary, uncertainty in powder particle properties, and many others \citep{moges2019review}. 

Conventional methods for simulating surface roughness and porosity rely on detailed powder scale simulations that account for irregular powder distributions and are restricted to  deterministic simulations of small regions. Kumar et al. proposed a mathematical model that includes both presence of particles on top surfaces and the stair step effect for surface roughness prediction \citep{STRANO2013589}, while a data fusion approach for surface roughness prediction is made by Wu et al. in 2018 \citep{10.1115/MSEC2018-6501}. Powder bed metal additive manufacturing (PBMAM) porosity relays on parameters and was presented with a physics-based model by Ning et al. \citep{ning2020analytical}. However, those models are computationally expensive and impede the consideration of part scale effects. Consequently, direct comparisons to engineering-size experimental measurements cannot be performed. Besides, the deterministic models with no stochastic information also restricts the prediction accuracy. 

In this work, we propose a stochastic modeling strategy to significantly accelerate the computational time for surface roughness and porosity and can be used for part-scale AM simulations. More specifically, an effective physics-based stochastic model is developed. It allows part scale simulations to quantify uncertainties in AM processes and predict defects such as surface roughness and porosity of the as-built parts. In particular, Higher-Order Proper Generalized Decomposition (HOPGD) \citep{2018IJNME.114.1438L,lu2019datadriven}, a specially designed tensor decomposition methods, is used as non-intrusive data learning and constructing reduced-order surrogate models for stochastic calibration. Experimental data \citep{schwalbach2021afrl}, including melt pool measurements, is used to stochastically calibrate the heat source module on AM-CFD software (developed in-house) \citep{11959300075f47619b6540a9764ebd85,gan2019benchmark}. The Markov Chain Monte Carlo (MCMC) sampling method is then used for the for statistical simulation predictions of process-structure-property that is capable of capturing the irregularities of LPBF scans. The proposed method has shown promise in predicting surface roughness and porosity for part scale simulations at very reduced computation costs while providing a high-fidelity computational model. Results for surface roughness, and porosity from this analysis tool will help provide highly accurate part scale property predictions that consider the inherent variation in melt pool geometry in AM processes. 
 
 The rest of the paper is organized as follows. Section 2 introduces  the framework of physics-based stochastic AM model. Section 3 briefly describes the experimental methods as well as the probability distributions of data. In Section 4, physics-based stochastic model is built. Section 5 shows validation of the stochastic model and predict the surface roughness and porosity.

\section{Proposed stochastic modeling framework for surface roughness and porosity prediction}
\label{sec:sample1}

The durability of AM parts has become an important topic to address for many applications, in which fatigue failure of materials has been one of the most common damage modes in cyclically loaded metallic materials. The fatigue life of AM parts is primarily governed by surface defects (e.g., surface roughness \citep{ZHAO201876})  and volumetric structural defects (e.g., porosity \citep{QU2022110454}). We develop in this work effective stochastic physics-based models that will allow part scale simulations to predict surface roughness and porosity of the as-built parts. The overall framework is shown in Figure 1. Experimental data, including melt pool dimension measured from AFRL experiment top-down and cross-section images, will be used to stochastically calibrate the heat source module on AM-CFD code for thermal fluid analysis. In the AM-CFD simulation, some heat source parameters show strong uncertainties and will be further calibrated based on the  experimental statistical distributions of melt pool data, where some statistical methods like the kernel density estimation \citep{Davis2011} and Kullback–Leibler divergence \citep{10.1214/aoms/1177729694} will be applied. To significantly reduce the computational cost for multi-parametric calibration, Higher-Order Proper Generalized Decomposition (HOPGD) \citep{2018IJNME.114.1438L,lu2019datadriven} is used to handle the model parameter calibration problem. The calibrated stochastic AM-CFD can then simulate part-scale samples using a Markov chain Monte Carlo (MCMC) method \citep{Hastings1970MonteCS} by sampling the calibrated heat source parameters in different time series, with results better than deterministic models. The AM-CFD will be capable of predicting the surface roughness and lack-of-fusion (LOF) porosity of the as-built parts by simulating multilayer-multitrack models.  

\begin{figure}[h!]
\centering
\includegraphics[width=1\textwidth]{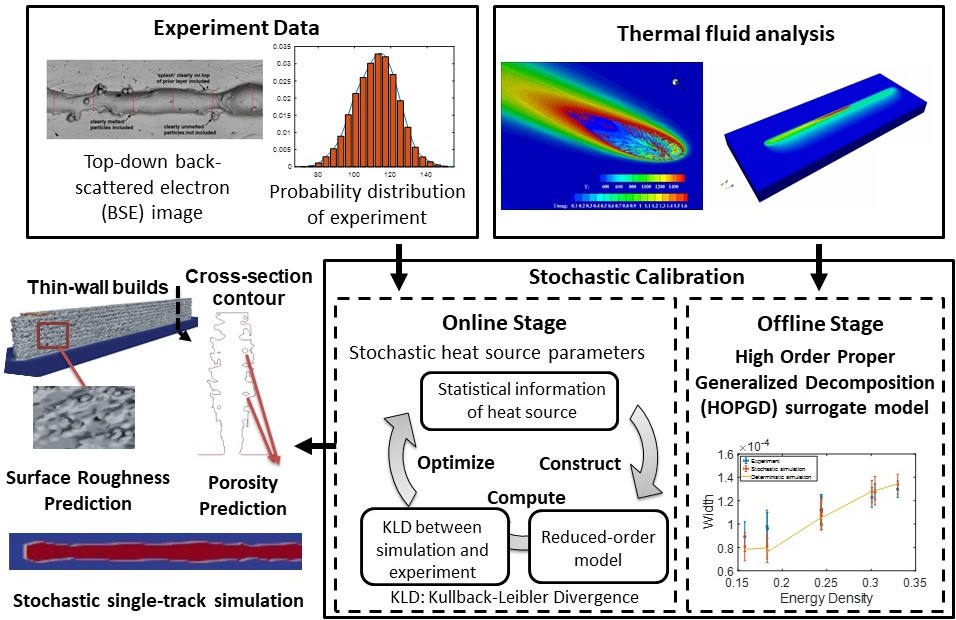}
\caption{Stochastic additive manufacturing simulation framework}
\end{figure}

\section{AFRL experimental data}
\label{sec:sample1}

\subsection{Stochastic expression of experimental data}
Laser powder bed fusion (LPBF) processes strongly trigger evaporation from the metal surface and the complex gas flow that disrupt the uniformity of the printed structure and subsequently result in properties of products. The computational model plays a significant role in understanding the process–structure–properties linkages in additive manufacturing (AM), and a well-designed experiment is required to ensure the accuracy of the models. In November 2019, the United States Airforce Research Laboratory: Materials and Manufacturing Directorate Structural Materials, Metals Branch (AFRL/RXCM) and America Made publicly announced the Additive Manufacturing Modeling Challenge Series, which provided a series of highly controlled additive manufacturing experiments for validation and quantification of computational models \citep{Cox2021AFRLAM}.

In the AFRL experiment, different cases including single-layer single-track, single-layer multi-track, and multi-layer single-track (thin-wall) builds of IN625 powder are produced with an EOS M280 commercial L-PBF system. To calibrate stochastic model, the single-track experiments with statistical measurements are used, while multi-track and multi-layer cases give validation to calibrated model and also measure surface roughness and lack-of-fusion porosity. Measurements of melt pool dimensions were taken using a combination of electron backscatter diffraction, to obtain the top-down description of the track (Fig.2a) and optical microscopy on etched cross sections (Fig.2b). The full description of the experimental setups and measurement procedures can be found in reference.

\begin{figure}[h!]
	\centering
	\subfigure[Top-down description and melt pool width measurements]{
		\begin{minipage}{8cm} %[b]%{0.2\textwidth} 
                       \includegraphics[width=\textwidth]{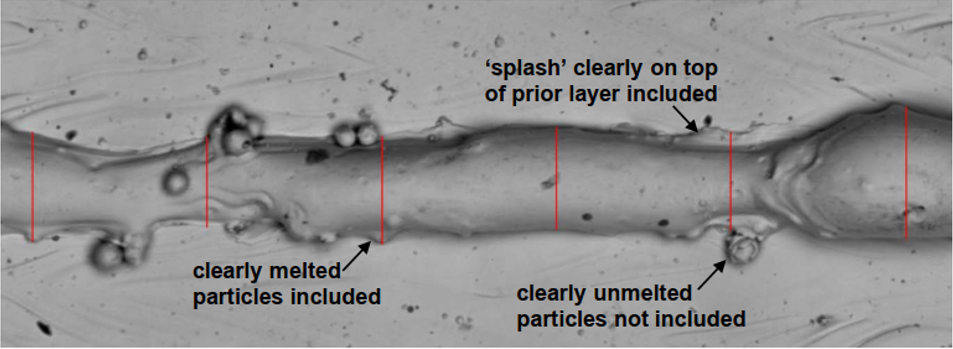}
		\end{minipage}
	}
    ~
	~	
	\subfigure[Top-down description and melt pool depth measurements]{
		\begin{minipage}{6cm}%[b]%{0.2\textwidth}
			\includegraphics[width=\textwidth]{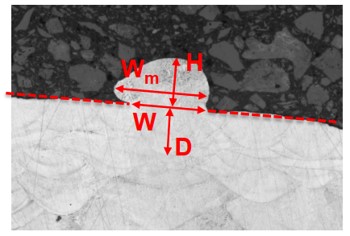} 
		\end{minipage}
		}
\caption{AFRL experiment measurements \citep{Cox2021AFRLAM}: a) Top-down and b) cross-section melt pool description. In the top-down description, the red lines (Fig.2 a)) are samples of melt pool width measurements. In cross-section descriptions, W is the width of melt pool, Wm is the largest value of all widths, D and H are the depth and height of the deepest position of melt pool.}	
\end{figure}

For identifying the best heat source parameters, it was necessary to know the influence of heat source parameters on the dimension. In particular,
the width (measured from top-down descriptions) and depth (the sum of D and H in cross-section images) of the single-track melt pool were the quantities of interest. Measured width and depth are shown in Table 1 and 2, where $\mu$ is the mean value and $\sigma$ is the standard deviation.

\begin{table}[htbp]
  \centering
  \caption{Width $(\mu m)$ measurement for 11 single-track cases (A11-A11)}
  \resizebox{\textwidth}{35mm}{
    \begin{tabular}{|c|c|c|c|c|c| p{3cm}|c|}
    \hline
    \multicolumn{1}{|p{3.89em}|}{Case } & \multicolumn{1}{p{4.61em}}{Laser } & \multicolumn{1}{|p{4.22em}|}{Scan} &       &       &       &  \\
    \multicolumn{1}{|p{3.89em}|}{number} & \multicolumn{1}{p{4.61em}}{Power } & \multicolumn{1}{|p{4.22em}|}{ Speed} & \multicolumn{1}{p{6em}|}{20 locations } & \multicolumn{1}{p{6em}|}{30 locations  } & \multicolumn{1}{p{6em}|}{40 locations} & \multicolumn{1}{p{6em}|}{\; 50 locations } \\
          & \multicolumn{1}{p{4.61em}}{($W$)} & \multicolumn{1}{|p{4.22em}|}{($mm/s$)} & $\mu$ \qquad $\sigma$ & $\mu$ \qquad $\sigma$ & $\mu$ \qquad $\sigma$ & \qquad $\mu$ \qquad $\sigma$\\
          \hline
    A1   & 300   & 1230  & 112.0±11.1 & 111.1±11.2 & 112.0±11.1 & \quad 111.9±10.9 \\
          \hline
    A2   & 300   & 1230  & 112.0±11.9 & 111.1±11.2 & 112.0±11.1 & \quad 111.9±10.9 \\
          \hline
    A3   & 290   & 953   & 127.6±7.0 & 124.7±9.1 & 125.5±10.5 & \quad 125.5±10.0 \\
          \hline
    A4   & 370   & 1230  & 122.9±8.4 & 119.4±10.0 & 117.7±10.2 & \quad 118.9±10.4 \\
          \hline
    A5   & 225   & 1230  & 96.0±13.9 & 100.1±14.2 & 99.7±13.1 & \quad 99.9±13.3 \\
          \hline
    A6   & 290   & 1588  & 97.9±14.0 & 99.7±11.3 & 100.7±13.6 & \quad 100.1±13.8 \\
          \hline
    A7   & 241   & 990   & 112.0±13.0 & 111.0±11.6 & 110.0±10.7 & \quad 109.4±10.5 \\
          \hline
    A8   & 349   & 1430  & 110.7±11.3 & 113.3±11.6 & 113.7±11.0 & \quad 113.4±11.3 \\
          \hline
    A9   & 300   & 1230  & 112.7±12.7 & 111.1±11.2 & 112.0±11.1 & \quad 111.9±10.9 \\
          \hline
    A10   & 349   & 1058  & 129.9±7.0 & 128.3±9.7 & 127.5±9.8 & \quad 127.3±9.4 \\
          \hline
    A11   & 241   & 1529  & 89.3±12.8 & 88.9±12.4 & 90.5±13.7 & \quad 90.8±13.4 \\
          \hline
    \end{tabular}%
  \label{tab:addlabel}}%
\end{table}%

% Table generated by Excel2LaTeX from sheet 'Sheet2'
\begin{table}[htbp]
  \centering
  \caption{Depth ($\mu$ m) measurement for 11 single-track cases (A1-A11)}
  \resizebox{\textwidth}{30mm}{
    \begin{tabular}{|c|c|c|c|c|c|c|p{1cm}|}
              \hline
    \multicolumn{1}{|p{4.055em}|}{Case } & \multicolumn{1}{p{4.055em}|}{Laser } & \multicolumn{1}{p{4.055em}|}{Scan} & \multicolumn{1}{p{6em}|}{Cross Section  } & \multicolumn{1}{p{6em}|}{Cross Section  } & \multicolumn{1}{p{6em}|}{Cross Section  } & \multicolumn{1}{p{6em}|}{Cross Section } &  \\
    \multicolumn{1}{|p{4.055em}|}{number} & \multicolumn{1}{p{4.055em}|}{Power } & \multicolumn{1}{p{4.055em}|}{ Speed} & \multicolumn{1}{p{7em}|}{Height (AFRL) } & \multicolumn{1}{p{8em}|}{Height (this work)  } & \multicolumn{1}{p{7em}|}{Depth (AFRL)  } & \multicolumn{1}{p{8em}|}{Depth (this work)  } & \multicolumn{1}{p{6em}|}{ Depth (sum) } \\
          & \multicolumn{1}{p{4.055em}|}{($W$)} & \multicolumn{1}{p{4.055em}|}{ ($mm/s$)} & $\mu$ \qquad $\sigma$ & $\mu$ \qquad $\sigma$ & $\mu$ \qquad $\sigma$ &  $\mu$ \qquad $\sigma$ & $\mu$   \hspace{0.6em}    $\sigma$ \\
                    \hline
    A1   & 300   & 1230  & 59.1±12.3 & 59.0±12.9 & 54.3±9.0 & 54.3±8.9 & 113.3±13.4 \\
              \hline
    A2   & 300   & 1230  & 65.7±21.8 & 65.7±21.7 & 52.3±9.0 & 52.5±8.6 & 118.2±19.9 \\
              \hline
    A3   & 290   & 953   & 68.1±9.2 & 68.1±9.1 & 72.0±7.4 & 72.0±7.4 & 140.0±12.8 \\
              \hline
    A4   & 370   & 1230  & 66.0±15.5 & 66.2±15.3 & 75.9±7.6 & 75.9±7.2 & 142.1±17.4 \\
              \hline
    A5   & 225   & 1230  & 60.3±14.9 & 60.3±14.9 & 25.0±6.1 & 25.0±6.1 & 85.3±13.6 \\
              \hline
    A6   & 290   & 1588  & 62.2±18.3 & 62.2±18.4 & 26.9±5.4 & 27.1±5.6 & 89.3±19.9 \\
              \hline
    A7   & 241   & 990   & 61.2±11.9 & 61.2±11.9 & 42.5±6.6 & 42.6±7.2 & 103.8±13.2 \\
              \hline
    A8   & 349   & 1430  & 60.1±15.9 & 60.1±16.1 & 58.5±4.6 & 58.5±4.6 & 118.5±18.2 \\
              \hline
    A9   & 300   & 1230  & 68.8±25.9 & 68.8±26.0 & 46.9±9.3 & 46.8±8.8 & 115.5±30.6 \\
              \hline
    A10   & 349   & 1058  & 63.5±17.8 & 63.3±17.6 & 84.0±8.9 & 83.8±8.6 & 147.1±19.4 \\
              \hline
    A11   & 241   & 1529  & 56.3±18.1 & 56.3±18.3 & 20.1±7.1 & 20.1±7.1 & 76.4±22.1 \\
              \hline
    \end{tabular}%
  \label{tab:addlabel}}%
\end{table}%

Table 1 indicates different location measurements of melt pool width, in which the forth column (20 locations) shows the result of AFRL measurement, while the fifth to seventh columns (30 to 50 locations) are measurements of this work. Similarly, comparisons of melt pool depth between AFRL and this work are shown in Table 2, while Depth (the last column) is the sum of cross-section depth and height. 

\subsection{Probabilistic model of the experimental data}

To further calibrate the heat source model, probability density functions (PDF) of experimental melt pool dimensions width and depth are required and calculated from the measurement sample using Kernel Density Estimation (KDE), which is a general and powerful way of estimating the probability density function of a random variable. A breif introduction of KDE is shown in the Appendix. The PDFs for are 11 single-track experiment cases are shown in Fig.3. 

\begin{figure}[h!]
\centering
\includegraphics[width=1\textwidth]{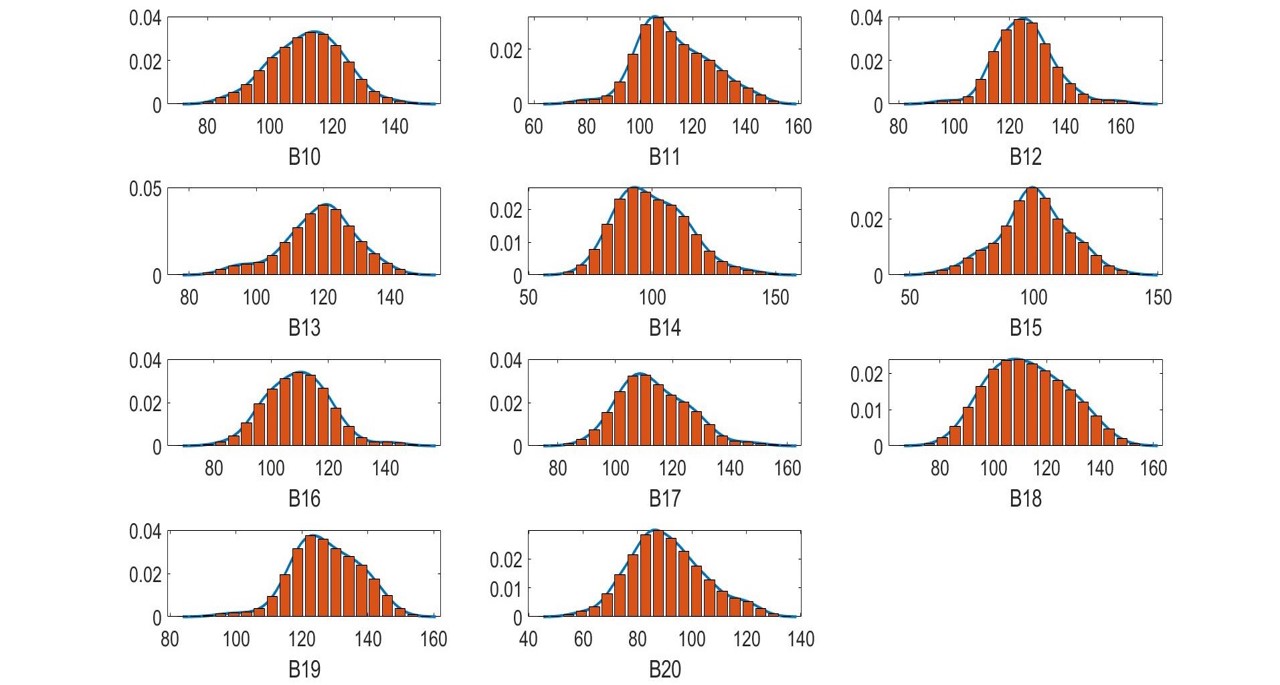}
\caption{Experimental measurement (orange bar) and its probability density function generated by KDE (blue line) of single experiment case A1 to A11. x axis represents width, while y axis is probability density.}
\end{figure}

Figure 3 shows that the PDF generated by KDE matches well with experiment measurements, which will be further used to calibrated stochastic heat source parameters in the next section.

\subsection{Basic Equations for thermal‑fluid additive manufacturing }

In this work, a thermal-fluid model considering liquid flow inside the melt pool driven by the Marangoni effect was developed as the AM process model to predict the melted track geometries of the L-PBF process. The basic theory of a well-tested transient three-dimensional thermal-fluid model that predicts the thermal field in the whole part and velocity field in the melt pool region is given. To solve the thermal-fluid model, the governing equations for mass, momentum, and energy conservation are given as follows: \citep{lu2020adaptive, 11959300075f47619b6540a9764ebd85}

\begin{equation}
\int_{\Omega f_l}{(\rho\nabla\cdot\bm{u})}dV=0
\end{equation}

\begin{equation}
\int_{\Omega f_l}\left(\frac{\partial(\rho \bm{u})}{\partial(t)}+\nabla\cdot{\rho \bm{u} \bm{u}}-\mu\nabla^2u+\nabla p+\frac{180\mu{(1-f_l)}^2}{c^2(f_l^3+B)}\bm{u}-\rho_{0} g\beta(T-T_{0}) \bm{I}\right)dV=0
\end{equation}

\begin{equation}
\int_{\Omega}\left(\frac{\partial(\rho h+\rho \Delta H)}{\partial t}+\nabla\cdot(\rho \bm{u} h+\rho \bm{u} \Delta H+\nabla\cdot \bm{q})\right)dV=0
\end{equation}

\noindent where $t$ is the time, $\bm{u}$ is the velocity, $\mu$ denotes the viscosity, $p$ is the pressure,  $T$ is the temperature, $\rho$ is the density, and $\beta$ is the thermal expansion coefficient. $g$ is the acceleration of gravity and equals to 9.8 $m/s^2$.  $\rho_{0}$ and $T_{0}$ are density and temperature of reference material. $H$ is the specific enthalpy, and can be divided into the sum of sensible heat $h$ and the
latent heat of fusion $\Delta H$. In this paper, $\mu$ is set as a constant, $c$ is the approximate primary dendritic spacing, which is set to 1 $\mu$m. $B$ is used to avoid division by zero and set as $10^{-6} m$. $f_l$ is the volume fraction of the liquid phase, which is defined as:

\begin{equation}
\begin{cases}                                   

                   f_l=0\ \ \ \ \ \ {if} \ \ \  T\leq T_s\\            

                   f_l=\frac{T-T_s}{T_l-T_s}\ \ \ \ \ \ {if} \ \ \  T_s \textless T \textless T_l\\
                   
                   f_l=1\ \ \ \ \ \ {if} \ \ \  T\geq T_l\\

                   \end{cases}
\end{equation}

\noindent where $T_s$ and $T_l$ are the solidus and liquidus temperature of materials, respectively.

Considering $\bar{\bm{q}}$ on the surface boundary, heat flex $\bm{q}$ and its relation with temperature $T$ is 

\begin{equation}
\bm{q}=-\bm{k}\cdot \nabla T
\end{equation}

\noindent where $\bm{k}$ is the thermal conductivity tensor. In isotropic cases, $\bm{k}=k\bm{I}$ denotes the second-order identity
tensor. The heat source and boundary condition can be written as:

\begin{equation}
\begin{cases}                                   

                   \bar{\bm{q}}\cdot \bm{n}=h_c(T-T_{0})-\sigma_s\varepsilon(T^4-T_{0}^4)+q_{source} \ \ \ \ \ \    on \ \ \partial\Omega_q\\            

                   T=\bar{T}\ \ \ \ \ \ \ \ \ \ \ \ \ \  \ \ \ \ on \ \ \partial\Omega_q\\             

                   \end{cases}
\end{equation}

\noindent where $h_c$ defines the convective heat transfer coefficient, $\sigma_s$ is the Stefan–Boltzmann constant, $\varepsilon$ is the emissivity, $\bm{n}$ is the normal direction of heat source surface.

The heat source $q_{source}$ from the laser, is described by a cylindrical shape with a Gaussian distribution described below:

\begin{equation}
q_{source}=\begin{cases}                                   

                   \frac{\epsilon Q \eta}{\pi r_b^2d}\exp{\left(\frac{-2\left(x_b^2+y_b^2\right)}{r_b^2}\right)}\ \ \ z_{top}-z\le\ d;\\            

                   0\ \ \ \ \  \ \ \ \ \ \ \ \ \ \ {\ \ \ \ \ \ \ \ \ \ \ \ \ \ z}_{top}-z \textgreater\ d\\             

                   \end{cases}
\end{equation}

 \noindent where $Q$ denotes the flux intensity,$\epsilon$ denotes the intensity factor, $\eta$ is the absorptivity, $r_b$ is the laser beam radius, $d$ is the depth of the heat source, and $z_{top}$ is the z-coordinate of the top surface of the computational domain. $x_b$ and $y_b$ are the coordinates in the local
reference system attached to the moving heat source. Note that the parameters, $\eta$, $r_b$ and $d$, are all unknown and uncertain variables, which are highly correlated to the vapor depression phenomenon in the L-PBF process. During calibration, the minimum value of absorptivity is limited to 0.28\citep{osti_1569671}. It is reported from the literature \citep{osti_1569671,2018JLasA..30c2410F} that as the laser power increases or the scan speed decreases, a vapor-induced depression appears and deepens, which leads to higher absorptivity caused by multiple reflections of the laser beam between the liquid and gas interface. Thus, we assume the three parameters, $\eta$, $r_b$ and $d$ are related to the ratio of laser power $P$ to scan speed $V$, shown as:
 
 \begin{equation}
 d=P1\frac{P}{V}
\end{equation}

 \begin{equation}
\eta=\max(P2\frac{P}{V},0.28)
\end{equation}

 \begin{equation}
 r_b=P3\frac{P}{V}
\end{equation}

$P1,P2,P3$ are three unknown parameters which require calibration using HOPGD method (described in section 4.1). The boundary condition for Eq.2 at the top surface is equal to the surface tension (i.e. Marangoni force):

 \begin{equation}
\tau_x=\mu\frac{\partial u_x}{\partial z}=\frac{d\gamma}{dT}\nabla_xT
\end{equation}

 \begin{equation}
\tau_y=\mu\frac{\partial u_y}{\partial z}=\frac{d\gamma}{dT}\nabla_yT
\end{equation}

\noindent where $\gamma$ is the surface tension, which depends on both temperature and materials, and $\frac{d\gamma}{dT}$ is the temperature
coefficient.

The powder layer is treated as a continuous media, and it is distinguished from the substrate through its material properties. A consolidated factor $\alpha$ ranging from 0 to 1 is used to identify the material state. The value of 0 stands for the material is in the original powder state (no consolidation), while 1 denotes a bulk state (fully consolidated). The definition of $\alpha$ is:

 \begin{equation}
\alpha=\frac{T_{peak}-T_s}{T_l-T_s}
\end{equation}

\noindent where $T_{peak}$ is the local peak energy, and $T_s$\ and $T_l$ are solid and liquid temperature of material, respectively.

The thermophysical properties of IN625 are summarized in Table 3. The densities at ambient and liquidus temperatures are used for solid and liquid densities, respectively. Temperature-dependent polynomials were used for the solid’s thermal conductivity and solid’s specific heat capacity.

\begin{table}[htbp]
  \centering
  \caption{Thermo-physical properties of IN625 and process constants \citep{Cox2021AFRLAM,Capriccioli2009FEPF,osti_5337885,valencia2013thermophysical}}
  \resizebox{\textwidth}{50mm}{
    \begin{tabular}{|p{13em}|c|p{12em}|c|}
    \hline
    Property/parameter & Value & Property/parameter & Value \\

    \hline
    Solid density ($kg \; m^{-3}$) & 8440  & Convection coefficient ($W \; m^{-1} \; K^{-1}$) & 10 \\

    \hline
    Liquid density ($kg \; m^{-3}$) & 7640  & Latent heat of fusion ($KJ \; kg^{-1} \; K^{-1}$) & \multicolumn{1}{c|}{290} \\

    \hline
    Powder density ($kg \; m^{-3}$) & 4330  & Dynamic viscosity ($Pa \; s$) & $7 × 10^{-3}$ \\

    \hline
    Solidus temperature ($K$) & 1563  & Thermal expansivity ($1/K$) & $5 × 10^{-5}$ \\

    \hline
    Liquidus temperature ($K$) & 1623  & Surface tension ($N \; m^{-1}$) & \multicolumn{1}{c|}{1.8} \\

    \hline
    Solid specific heat capacity ($J \; kg^{-1} \; K^{-1}$) & \multicolumn{1}{p{8em}|}{0.2441$T$ + 338.39} & Marangoni coefficient ($N \; m^{-1} \; K^{-1}$) & -$3.8 × 10^{-4}$ \\

    \hline
    Liquid specific heat capacity ($J \; kg^{-1} \; K^{-1}$) & 709.25 & Emissivity & \multicolumn{1}{c|}{0.4} \\

    \hline
    Powder specific heat capacity ($J \; kg^{-1} \; K^{-1}$) & \multicolumn{1}{p{8em}|}{0.2508$T$ + 357.70} & Ambient temperature ($K$) & \multicolumn{1}{c|}{295} \\

    \hline
    Solid thermal conductivity ($W \; m^{-1} \; K^{-1}$) & \multicolumn{1}{p{8em}|}{0.0163$T$ + 4.5847} & Reference temperature ($K$) & \multicolumn{1}{c|}{295} \\

    \hline
    Liquid thermal conductivity ($W \; m^{-1} \; K^{-1}$) & 30.078 & Preheat temperature ($K$) & \multicolumn{1}{c|}{353} \\

    \hline
    Powder thermal conductivity ($W \; m^{-1} \; K^{-1}$) & 0.995 & Stefan–Boltzmann constant ($W \; mm^{-2} \; K^{-4}$) & $5.67 × 10^{-14}$ \\
    \hline

    \end{tabular}%
  \label{tab:addlabel}}%
\end{table}%

In order to consider the influence of the localized preheating from adjacent scan paths that leads to transient behavior of the vapor depression, the residual heat factor (RHF) is considered into the heat source model \citep{yeung2020residual}. RHF at specific point $i$ is defined as:

 \begin{equation}
{RHF}_i=\sum_{k\in S_i}{{(\frac{R-d_{ik}}{R})}^2\left(\frac{T-t_{ik}}{T}\right)}L_k
\end{equation}

The scan path is composed of discrete points defined by the time step of the simulation and the laser scan speed. $d_{ik}$ denotes the distance distance between point $i$ and $k$, which indicates the preheating on point i by a previously scanned point k. Similarly, elapsed time since $k$ was scanned is denoted by $t_{ik}$. $L_k$, the normalized laser power at point k, is equal to 1 when the laser is on, while 0 denotes the laser is off. $R$ and $T$ are constants with the values of $2\ast{10}^{-4}$ and $2\ast{10}^{-3}$, respectively. They play as thresholds that ignore points which had not interacted with the laser for a sufficient amount of time, while the other points within the threshold belong to set $S_i$, where $S_i=\{t_{ik}<T\cup\ d_{ik}<R,\ \ where\ i>k\}$. RHF is normalized as $RHF=\frac{{RHF}_i}{{RHF}_c}$, where ${RHF}_c$ equals to ${RHF}_i$ at the middle part of the toolpath, and it is greater than 1 at the corner of the toolpath, as shown in the fllowing figure. 

\begin{figure}[h!]
\centering
\includegraphics[width=0.5\textwidth]{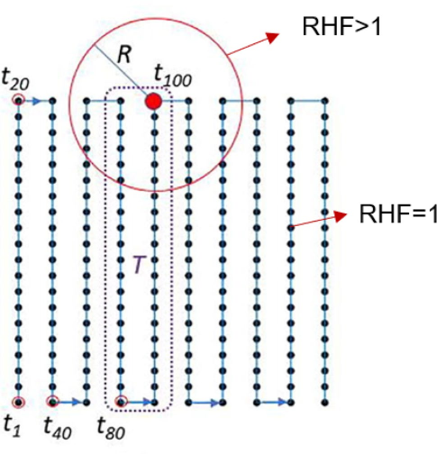}
\caption{Quantitative comparisons of cross-sectional area between experimental measurements and stochastic simulation \citep{yeung2020residual}}
\end{figure}

Considering the influnence of residual heat, the heat source parameters can be coupled with the RHF as:

 \begin{equation}
 d=P1\frac{P}{V}{\rm RHF}^2
\end{equation}

 \begin{equation}
\eta=\max(P2\frac{P}{V}{\rm RHF}^2,0.28)
\end{equation}

 \begin{equation}
 r_b=P3\frac{P}{V}{\rm RHF}^2
\end{equation}

where three unknown parameters $P1,P2,P3$ will be further calibraetd with HOPGD method.

\section{Stochastic calibration for process parameters}

The Higher-Order Proper Generalized Decomposition (HOPGD) based data-driven reduced order model is used to do stochastic calibration for process parameters to find the interaction between the metal and the laser heat source.

\subsection{Tensor decomposition based surrogate model}

One challenge for accurate AM simulations is the identification of model parameters.  One standard way for unknown parameters calibration is the genetic algorithm approach, in which evaluation of parameters requires repeated calls of the computational model.  This results in high computational costs, and thus, a HOPGD-based reduced order surrogate model is used to accelerate the parameter calibration (identification) process significantly.

HOPGD, a non-intrusive data learning and constructing reduced order surrogate models, relies on the database and falls into the family of data-driven approaches. The database can be either from simulations or experiments, and the foundation behind tensor decomposition is the separation of variables technique. For a n-dimensional function $f\left(\mu_1,\mu_2,\ \mu_{3\ \ldots.,\ }\mu_n\right)$ that contains the quantity of interest as a function of n parameters, HOPGD separation form is:

 \begin{equation}
f(\mu_1,\mu_2, \mu_3 \ldots.,\mu_n)\approx\sum_{m=1}^{k}{{F}_1^{(m)}(\mu_1){F}_2^{(m)}(\mu_2)}{F}_3^{(m)}(\mu_3)\ldots.,{F}_n^{(m)}(\mu_n)
\end{equation}

The function $f$ is given by the finite sum of products of the separated functions ${F}_{i}^{\left(m\right)}(i=1,..,n)$. ${F}_{i}^{\left(m\right)}$ identifies the variation of function ${f}$ in the parameter direction $\mu_i$, which is also called mode function. $n$ is the rank of approximation and $m$ defines the mode number of each component (not represent exponential terms). The number of $n$ is priori unknown and can be obtained with a precomputed physics-based simulation database \citep{2018IJNME.114.1438L,lu2019datadriven,lu2018multi,lu2018space}.  HOPGD seeks the projection of data for computing the mode functions that can reproduce the original function, and thus can be used as a surrogate model for fast prediction. Examples of codes can be found on the GitHub project (https://yelu-git.github.io/hopgd/).

In our AM computational model, we seek to identify the relations between heat source model and important parametric melt pool dimensions (width and depth), unknown parameters are calibrated with the following HOPGD model:

 \begin{equation}
{Y}_{s}=\left[\begin{matrix}W_{s}\\D_s\\\end{matrix}\right]\approx\left[\begin{matrix}W_{PGD}\\D_{PGD}\\\end{matrix}\right]={F}\left(e,P1,\ P2,\ P3\right)=\sum_{m=1}^{k}{{F}_1^{\left(m\right)}\left(e\right){F}_2^{\left(m\right)}\left(P1\right){F}_3^{\left(m\right)}\left(P2\right)}{F}_4^{\left(m\right)}\left(P3\right) \label{eq13}
\end{equation}

\noindent where $P1,\ P2,\ P3$ are heat souse parameters, $e=\frac{P}{V}$ is the energy density. Note that all parameters in this section are deterministic. $W_{s}$ and $D_s$ are melt pool dimensions width and depth calculated from simulation. $W_{PGD}$ and $D_{PGD}$ are predictions of width and depth from the HOPGD surrogate model. Once the HOPGD surrogate model is constructed, real time predictions can be used in an optimization problem to find calibrated parameters ${p}^\ast$. The optimization problem for the calibration can be written as

 \begin{equation}
p^\ast=arg\ min\left[\ J(W_{PGD},W_{e},{p})+J(D_{PGD},D_{e},{p})\ \right]\label{eq14}
\end{equation}

\noindent where $W_e$ and $D_e$ are AFRL single-track width and depth measurements. ${{p}}=[P1,P2,P3]$, and $J$ denotes the objective function that measures the distance between the surrogate model’s predictions and the experimental measurements. The above surrogate model used significantly reduce the computational cost, since only 1D interpolation is required to find output of a given point after finding functions ${F}_{i}^{\left(m\right)}$. The method is applicable to deal with high-dimensional problems with limited costs, which is challenging for other methods.

The steps to solve the optimization problem in Eq.\ref{eq14} are:

\begin{enumerate}
    \item 	Sample the parameter space with the adaptive
sparse grid strategy \citep{lu2019datadriven,saha2021microscale} and compute the simulated melt pool dimensions $(W_{s}, D_s)$ with the AM-CFD model for the selected data points.
    \item 	Construct HOPGD surrogate model and calculate $W_{PGD}$ and $D_{PGD}$ with Eq.\ref{eq13} for sample data.
    \item   Solve the optimization problem in Eq.\ref{eq14} with HOPGD surrogate model to calibrate heat source parameters.
\end{enumerate}

\subsection{Stochastic calibration for heat source parameters}

Section 4.1 shows the framework of building a surrogate model and a deterministic strategy to calibrate process parameters. To further indicate surface roughness and porosity in the L-PBF process, stochastic heat source parameters which contains noise information are required to be calibrated. The stochastic calibration can be based on the same HOPGD model presented previously. 

This time, the stochastic heat source parameters are assumed to satisfy a tri-variate normal distribution:

 \begin{equation}
{{p}}\left(P1,P2,P3\right)=\frac{1}{{(2\pi)}^\frac{3}{2}\left|{\Sigma}\right|^\frac{1}{2}}e^{-\frac{1}{2}\left[\left({P}-{\mu}\right)^T{\Sigma}^{-1}\left({P}-{\mu}\right)\right]}
\end{equation}

\noindent where ${{P}}=[P1,P2,P3]^ { T }$ is the vector of heat parameters, $\mu=[\mu 1,\mu 2,\mu 3]^ { T }$ is the mean vector. ${\Sigma=\left[\begin {matrix}C11&C12&C13\\C21&C22&C23\\C31&C32&C33\\ \end{matrix}\right]}$ is the covariance matrix. Due to the symmetry of $\Sigma$, unknown coefficients are ${\mu_1,\mu_2,\mu_3,C}_{11},C_{22},C_{33},C_{12},C_{23},C_{13}$ whose distributions are required to be calibrated. The calibrated stochastic AM-CFD can then simulate part-scale samples using Markov chain Monte Carlo method. The stochastic AM-CFD will be capable of predicting the surface roughness and lack-of-fusion (LOF) porosity of the as-built parts by simulating multilayer-multitrack models.

Similar to the processes of finding the distribution of experimental $W_{e}$ and $D_{e}$ ($f_{We}(x)$ and $f_{De}(x)$), the distributions of $W_{PGD}$ and $D_{PGD}$ ($f_{WPGD}(x)$ and $f_{PGD}(x)$) are also estimated with kernel density estimation. An optimization problem is defined to calibrate the stochastic parameters:

 \begin{equation}
{p}^\ast=arg\ min\left[\ J(W_{PGD},W_{e},{p})+J(D_{PGD},D_{e},{p})\ \right] \label{eq16}
\end{equation}

$W_{e}$ and $D_{e}$ are statistical experimental measurements that with mean and variance. Besides, $W_{PGD}$ and $D_{PGD}$ in Eq.\ref{eq16} also satisfy statistic distributions which are calculated based on stochastic parameters P1,P2,P3 in Eq.15. To define the distance $J$ between different melt pool distributions, Kullback-Leibler Divergence (KLD) \citep{10.1214/aoms/1177729694} will be introduced, which is required in the optimization of calibrated parameters. A brief introduction of Kullback-Leibler Divergence is in Appendix.

 KLD enables to qualify the distance between two distributions. If the KLD reaches the minimum, the probability density estimation expression can be considered to achieve its best estimation result. Thus, the objective function in Eq.\ref{eq16} can further be defined with KLD: 

 \begin{equation}
 \begin{aligned}
{p}^\ast=arg\min
\sum_{i=1}^{11}{f_{We\left(i\right)}\left(We\right)\log{\frac{f_{We\left(i\right)}\left(We\right)}{f_{WPGD\left(i\right)}\left(W_{PGD}(P1,P2,P3)\right)}}}  +\\ \sum_{i=1}^{11}{f_{De\left(i\right)}\left(De\right)\log{\frac{f_{De\left(i\right)}\left(De\right)}{f_{DPGD\left(i\right)}\left(D_{PGD}(P1,P2,P3)\right)}}}
\end{aligned}
\end{equation}

\noindent where i is the index of single-track cases. $f_{We}, f_{De},f_{WPGD}, f_{DPGD}$ are distributions of experimental width, experimental depth, simulated width and depth with HOPGD surrogate model, respectively. 

The distributions of experiment and simulation can be obtained through KDE \citep{Davis2011}:

 \begin{equation}
f_{WPGD(i)}(W)=\frac{1}{nh}\sum_{j=1}^{n}{K(\frac{W-W_{PGDj}}{h})}
\end{equation}

 \begin{equation}
f_{DPGD(i)}(D)=\frac{1}{nh}\sum_{j=1}^{n}{K(\frac{D-D_{PGDj}}{h})}
\end{equation}

 \begin{equation}
f_{We(i)}(W)=\frac{1}{nh}\sum_{j=1}^{n}{K(\frac{W-W_{ej}}{h})}
\end{equation}

 \begin{equation}
f_{De(i)}(D)=\frac{1}{nh}\sum_{j=1}^{n}{K(\frac{D-D_{ej}}{h})}
\end{equation}

\noindent where $k$ is the Gaussian kernel. $n$ is the number of sample points and $h$ is the bandwidth. Detailed expression are shown in the appendix.

\section{Validated melt-pool geometry prediction against AFRL experimental data}

In this section, different single-track, multi-track and multi-layer cases are validated with AFRL experimental data using the stochastic AM simulation model.

\subsection{Stochastic prediction of single-track melt pool}

The stochastic simulation model in the previous section enables to predict uncertainty of LBPF melt pool with stochastic process parameters and heat source parameters. To predict stochastic single-track melt pool,Markov Chain Monte Carlo (MCMC) \citep{Hastings1970MonteCS}, an algorithms for sampling from probability distributions based on time series, is used for samples generation and statistical simulation predictions of process-structure-property.

In each fluid dynamics time step, MCMC-sampled heat sources is imported into AM-CFD program, which helps for predicting surface roughness and porosity for part scale simulations at very reduced computation costs while providing a high-fidelity computational model. The comparisons between experiment, deterministic simulation (with constant heat source model), and stochastic simulation (with calibrated stochastic heat source model) is shown below.

The comparisons of overall geometry (Figure 5) and cross-section view (Figure 6) between deterministic and stochastic simulation is shown. Clear uncertain information is revealed in stochastic AM simulation. To verify the accuracy of stochastic simulation, statistical information (e.g. mean and variance for melt pool width and depth) between stochastic simulation, and experiment \citep{Cox2021AFRLAM} are summarized in Figure 6 and 7.    

\begin{figure}[h!]
\centering
\includegraphics[width=1\textwidth]{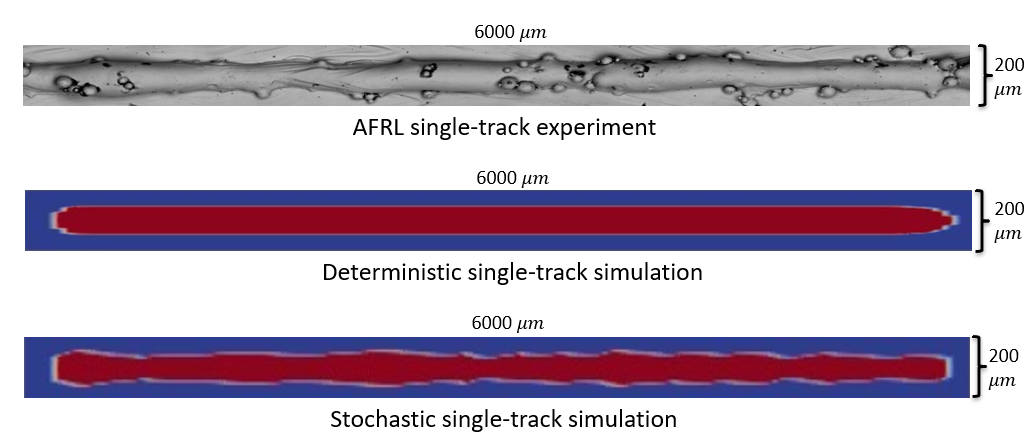}
\caption{Comparisons between AFRL experiment \citep{Cox2021AFRLAM}, deterministic simulation (with constant heat source model), and stochastic simulation (with calibrated stochastic heat source model) for single track}
\end{figure}

\begin{figure}[h!]
\centering
\includegraphics[width=1\textwidth]{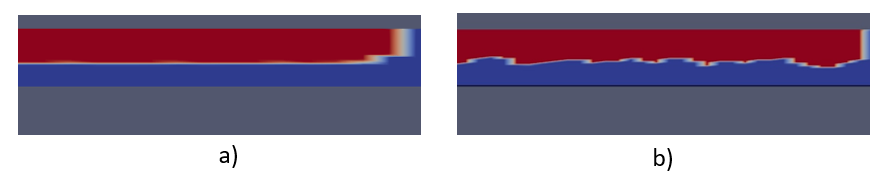}
\caption{Cross-section view comparisons between a) deterministic simulation and b) stochastic simulation for single track}
\end{figure}

\begin{figure}[h!]
\centering
\includegraphics[width=0.8\textwidth]{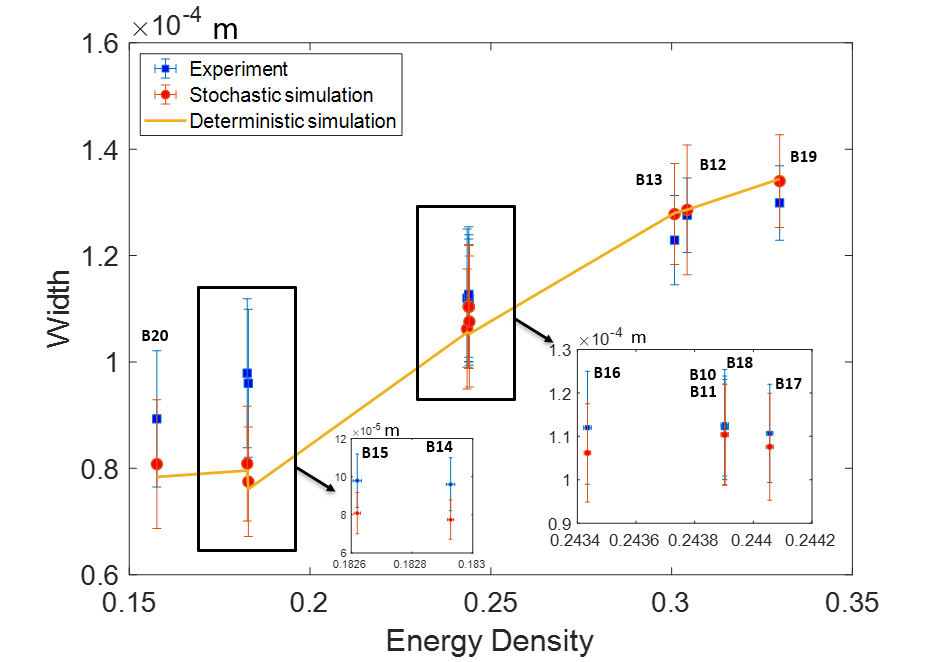}
\caption{Statistical information of melt pool width between stochastic simulation, deterministic simulation and experiment}
\end{figure}

\begin{figure}[h!]
\centering
\includegraphics[width=0.8\textwidth]{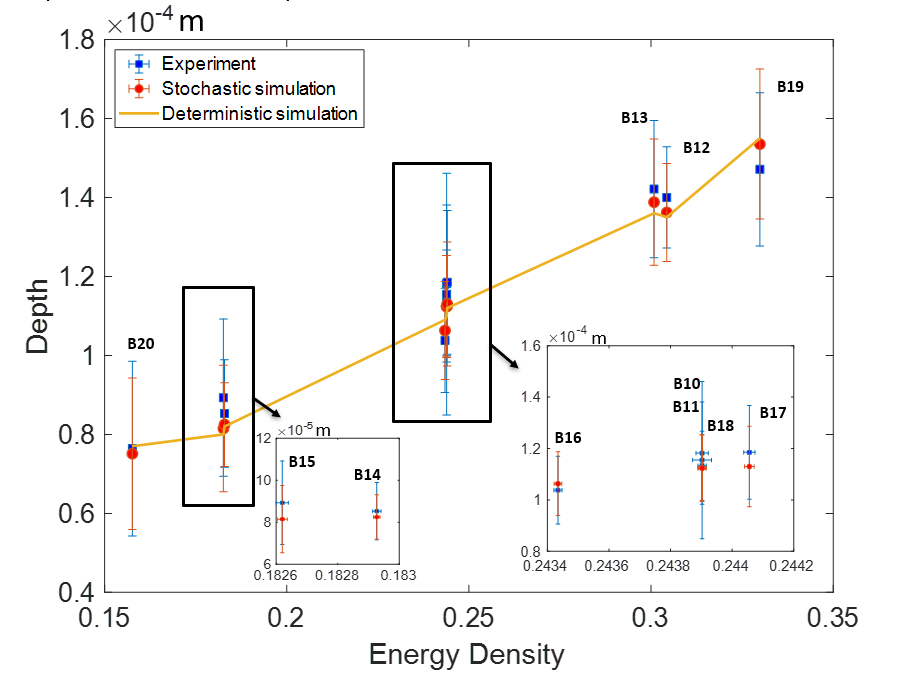}
\caption{Statistical information of melt pool depth between stochastic simulation, deterministic simulation and experiment}
\end{figure}

In figure 6 and 7, energy density for 11 different single-track cases are shown in x coordinates, while y coordinates present the melt pool width and depth, respectively. The blue and red error bars show mean and variance of experiment and stochastic simulation. The yellow line passes through the deterministic simulation melt pool dimensions. Cases with close energy density are zoomed in for detailed comparison. To calculate the mean and variance of melt pool dimensions, 50 width locations and 20 depth locations are measured from experiments and stochastic simulations. It can be seen from figure 6 and 7 that the stochastic simulation matches well experiment melt pool dimensions. Compared with deterministic simulation, stochastic simulation contains uncertain information of melt pool and show more accurate predictions for most cases.   

\subsection{Validation of multilayer and multi-track cases}

To further predict surface roughness and porosity, multi-layer (thin-wall) and multi-track cases are also simulated with stochastic AM-CFD model.

The simulation of two thin-wall specimens, B1 and B2, consists of 10 consecutive 40 $\mu m$ thick layers each with a unidirectional scanning track length of 5 mm \citep{Cox2021AFRLAM}. One of the thin walls used a laser power of 300 W and a scan speed of 1230 mm/s, whereas the
other used 241 W and 1529 mm/s as shown in Table 4. Figure 8a shows the simulated result for case B1. For a quantitative comparison, the wall is divided into three measurement zones shown in Figure 8a. The average and standard deviation of the height above the substrate pad datum and the total cross-sectional area for the entire portion of the wall above the substrate pad datum were measured for each measurement zone as shown in Fig. 8b. There is a minimum of 3 cross sections collected within Zones 1 and 3, and approximately 20 cross sections in Zone 2. Figure 9 shows the comparisons of the cross sectional area for the three different zones between the experimentally measured and computationally predicted for B1 and B2 multi-track cases. The simulated height and area match well with the measurements in the second and third zones, which indicates the developed model can predict the steady-state melt pool geometry well. However, at Zone 1, the beginning
region of each layer, the model underestimates the results.
This implies that some transient behaviors occurring at the
beginning of each layer are being neglected by the model.

\begin{table}[]
\caption{Multi-layer simulation process parameters \citep{Cox2021AFRLAM}}
\resizebox{\textwidth}{12mm}{
\begin{tabular}{|c|c|c|c|c|c|}
\hline
\begin{tabular}[c]{@{}c@{}}Case \\ Number\end{tabular} & \begin{tabular}[c]{@{}c@{}}Laser \\ Power (W)\end{tabular} & \begin{tabular}[c]{@{}c@{}}Scan \\ Speed (mm/s)\end{tabular} & \begin{tabular}[c]{@{}c@{}}Layer \\ thickness (\textbackslash{}mum)\end{tabular} & \begin{tabular}[c]{@{}c@{}}Track \\ length (mm)\end{tabular} & \begin{tabular}[c]{@{}c@{}}The number \\ of layers\end{tabular} \\ \hline
B1                                                    & 300                                                        & 1230                                                         & 40                                                                               & 5                                                            & 10                                                              \\ \hline
B2                                                    & 241                                                        & 1529                                                         & 40                                                                               & 5                                                            & 10                                                              \\ \hline
\end{tabular}
}
\end{table}

\begin{figure}[h!]
\centering
\includegraphics[width=1\textwidth]{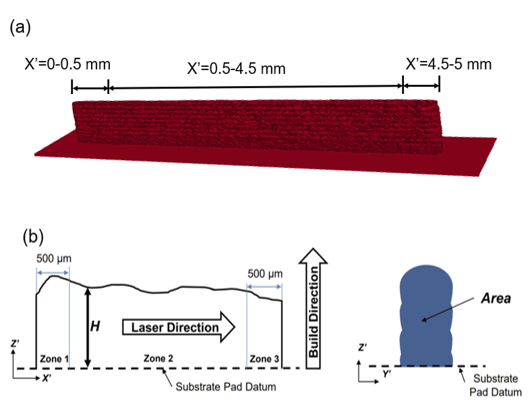}
\caption{As-built multi-layer structure and its measurements for case B1. a) Multi-layer simulation. b) A schematic of the cross section area measurements for three Zones \citep{Cox2021AFRLAM}}
\end{figure}

\begin{figure}[h!]
\centering
\includegraphics[width=1\textwidth]{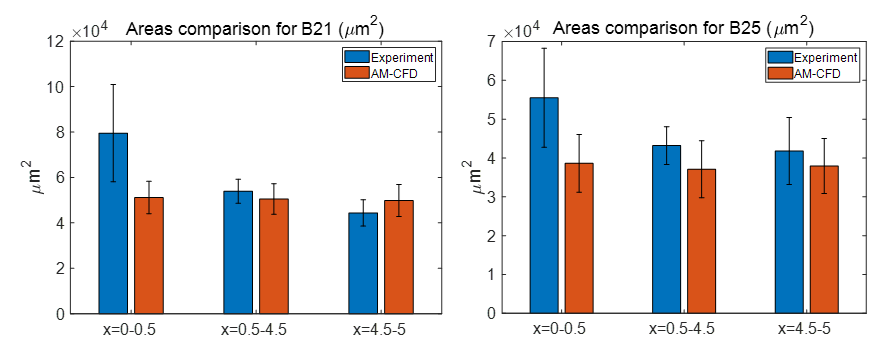}
\caption{Quantitative comparisons of cross-sectional area between experimental measurements and stochastic simulation}
\end{figure}

Six multi-track cases are also simulated with calibrated stochastic AM-CFD model to predict geometrical details
of the melted tracks for the L-PBF process. Figure 10 indicates the substrate geometries and tool paths of those six simulations (multi-track cases C1, C2, C3, C4, C5, and C6). The tool paths are labeled according to the L-PBF experiments performed by AFRL. A dwell time of 0.5ms is set in which the laser beam is turned off while the beam moves to the beginning position of the next scan path. The black frames show the substrate dimensions, and the arrows represent the laser scan paths. Table 5 summaries the process parameters used for all 6 multi-track cases. Quantitative comparisons of melted track geometries between experiment, deterministic and stochastic at the middle of the toolpath (x=1.5mm) for multi-track simulation C1 is shown in Figure 11 a) . The average and standard deviation of each quantity for the tracks are plotted and W defines the melt pool width and D stands for the melt pool depth, which are defined in Figure 11 b). The multi-track simulations matches well with experimental data, and have shown promise in high-precision AM predictions with incapability of capturing the variation in melt pool, which can further give potentials in predicting surface roughness and porosity for part scale simulations at very reduced computation costs.

\begin{figure}[h!]
\centering
\includegraphics[width=0.9\textwidth]{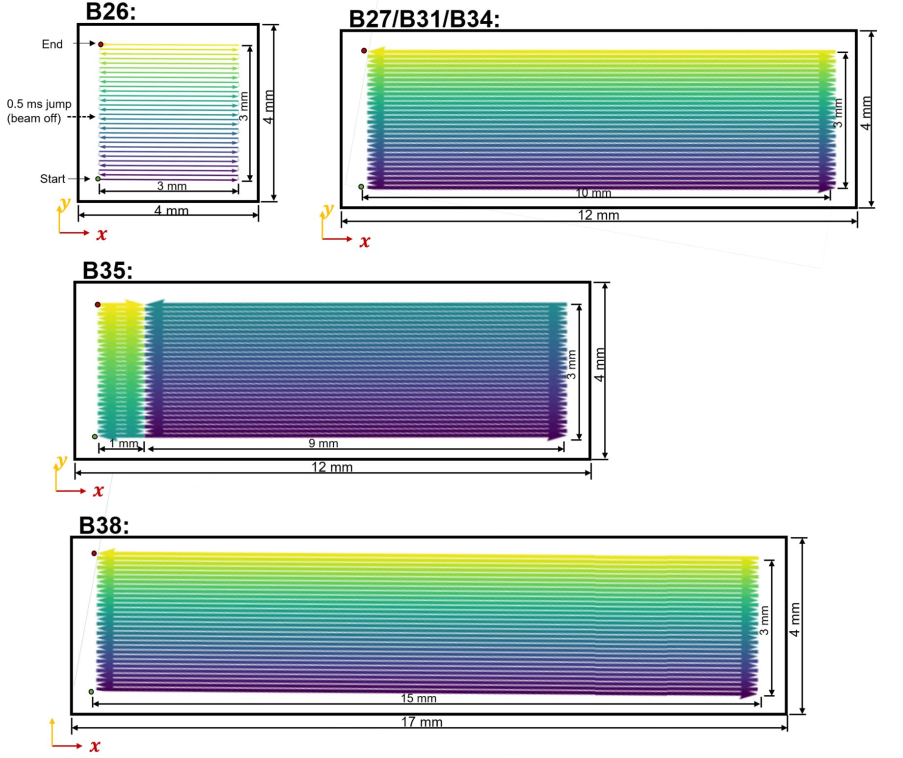}
\caption{Surface roughness and porosity for multi-layer simulation \citep{Cox2021AFRLAM}}
\end{figure}

\begin{figure}[h!]
\centering
\includegraphics[width=0.9\textwidth]{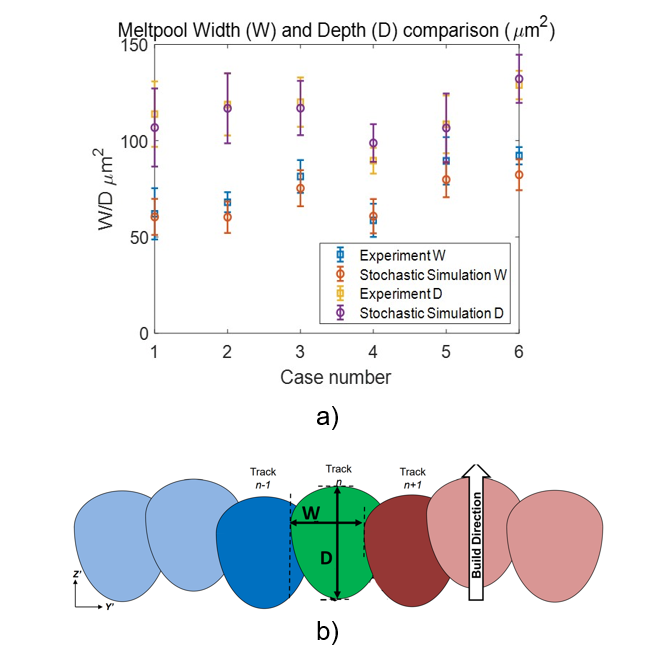}
\caption{Quantitative comparisons of melt pool dimensions between experimental and stochastic simulations, a) W and D are melt pool width depth, respectively. The average and standard deviation of each quantity for different tracks are plotted (The error bar represents the standard deviation.)}
\end{figure}

\begin{table}[]
\caption{Multi-track simulation process parameters \citep{Cox2021AFRLAM}}
\resizebox{\textwidth}{20mm}{
\begin{tabular}{|c|c|c|c|c|c|}
\hline
\begin{tabular}[c]{@{}c@{}}Case \\ Number\end{tabular} & \begin{tabular}[c]{@{}c@{}}Laser \\ Power (W)\end{tabular} & \begin{tabular}[c]{@{}c@{}}Scan \\ Speed (mm/s)\end{tabular} & \begin{tabular}[c]{@{}c@{}}Hatch \\ Spacing (mm)\end{tabular} & \begin{tabular}[c]{@{}c@{}}Toolpath plane \\ dimensions (mm)\end{tabular} & \begin{tabular}[c]{@{}c@{}}The number \\ of tracks\end{tabular} \\ \hline
C1                                                    & 300                                                        & 1230                                                         & 0.1                                                           & 3*3                                                                       & 30                                                              \\ \hline
C2                                                    & 300                                                        & 1230                                                         & 0.1                                                           & 10*3                                                                      & 30                                                              \\ \hline
C3                                                    & 300                                                        & 1230                                                         & 0.075                                                         & 10*3                                                                      & 40                                                              \\ \hline
C4                                                    & 300                                                        & 1230                                                         & 0.125                                                         & 10*3                                                                      & 24                                                              \\ \hline
C5                                                    & 300                                                        & 1230                                                         & 0.1                                                           & 10*3                                                                      & 30                                                              \\ \hline
C6                                                    & 290                                                        & 953                                                          & 0.1                                                           & 15*3                                                                      & 30                                                              \\ \hline
\end{tabular}
}
\end{table}

\section{Stochastic thermal‑fluid additive manufacturing simulation based surface roughness and porosity prediction}
\label{sec:sample1}
Fatigue failure plays an significant role in the durability of AM parts. The fatigue life of AM parts is primarily governed by surface defects (e.g., surface roughness) and volumetric structural defects (e.g., porosity). However, generally, the simulation of surface roughness or porosity are very difficult because of the lack of simulation model which is able to describe uncertain information accurately in statistical level but still simple enough to satisfy the efficiency requirement at the same time. Thanks to the validated stochastic physics-based AM-CFD model shown in the previous sections, we are capable of predicting the surface roughness and lack-of-fusion (LOF) porosity of the as-built parts by simulating multilayer-multitrack models, as shown below. To generate time-dependant sequences, Markov chain Monte Carlo method is used for part-scale sampling of the calibrated stochastic AM-CFD model.

\subsection{Prediction of surface roughness}

Figure 12 shows the multi-layer simulation which clearly indicates the surface roughness and porosity. The rest of this section shows how to measure the surface roughness and compared it with experimental results.

\begin{figure}[h!]
\centering
\includegraphics[width=1\textwidth]{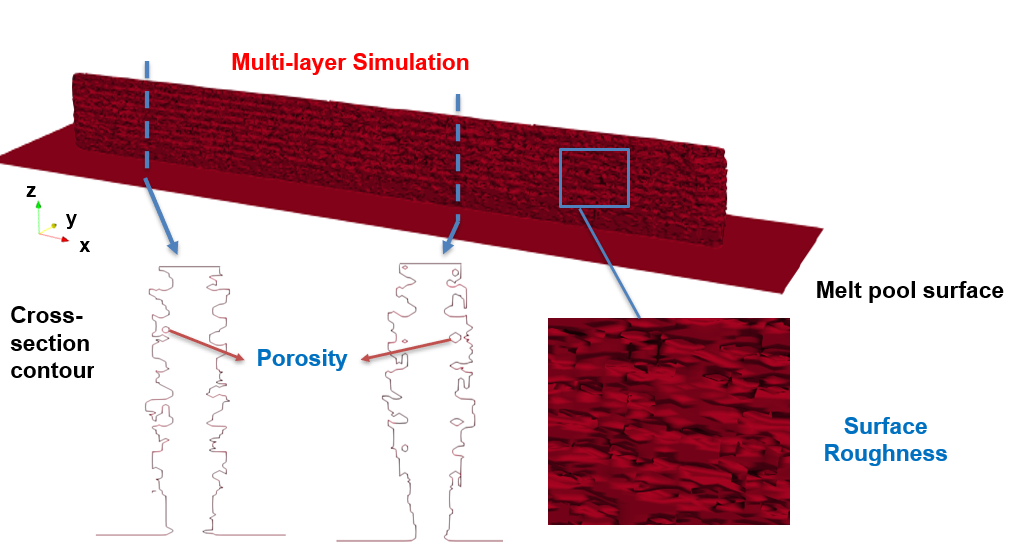}
\caption{Surface roughness and porosity for multi-layer simulation}
\end{figure}

The main roughness parameters reported is average roughness (Ra). It evaluate the average standard deviation of the heights (valleys and peaks) in a surface profile to compute the degree of roughness. For the computation of these parameters, it is first necessary to compute the fitting plane for the points acquired from the surface. From the plane coefficients, it is possible to determine the height of a peak or valley by evaluating the height coordinate of each point of the cloud. 

The equation of average roughness Ra is given by \citep{degarmo1997materials}:

 \begin{equation}
Ra=\frac{1}{A}\iint_{S}\left|f(y)\right|dS
\end{equation}

\noindent where $A$ is the sampling area and $f(y)$ is height of the profile. The simulated wall is equally divided into 10 regions, and the mean value and variance of multilayer cases B1 and B2 cases are thus calculated by considering different regions (B1:$Ra=12.62 \pm 2.21 \mu m$, B3:$Ra=14.57 \pm 2.78 \mu m$)

Surface roughness experiments \citep{koutiri2018influence} are used to validate the simulated surface roughness. In the experiment, the same material as AFRL Challenge2 is used (IN625), and the relation between surface roughness and volumetric energy density (VED) \citep{bertoli2017limitationsr} is given. Different from linear energy density $e$ in the previous section, the equation for VED is:

 \begin{equation}
VED=\frac{P}{V\sigma t}
\end{equation}

\noindent where $P$ is laser power, $V$ is scan speed, $\sigma$ is the laser beam diameter, and $t$ is the thickness for a single layer. VEDs for B1 and B2 multi-layer cases are in the following table, and the comparisons between experiment and simulation surface roughness are in Figure 11.

\begin{figure}[h!]
\centering
\includegraphics[width=0.8\textwidth]{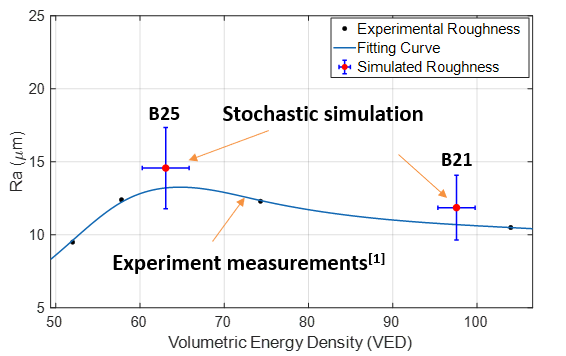}
\caption{Surface roughness and porosity for multi-layer simulation}
\end{figure}

In figure 13, the black dots are the experiment roughness with different VEDs. The blue curve is the fitting curve of experiment surface roughness. The two error bars are the simulated roughness with corresponding mean and variance for B1 and B2. The comparison shows the simulated roughness match well with experiment, which validate the accuracy of the stochastic AM simulation.

\section{Conclusion}
\label{sec:sample1}
In this paper, an effective phynsics-based stochastic modeling framework is proposed for LPBF simulations. Non-intrusive data learning method, including Higher-Order Proper Generalized Decomposition (HOPGD), is used for constructing reduced-order surrogate models for stochastic calibration. The Markov Chain Monte Carlo (MCMC) sampling method is then used for the for statistical simulation predictions of process-structure-property that is capable of capturing the irregularities of LPBF scans. The stochastic simulation predictions are validated against AFRL multi-layer and multitrack experiments. The results are more accurate when compared with regular deterministic simulations. The proposed method has shown promise in predicting surface roughness and porosity for part scale simulations at very reduced computation costs while providing a high-fidelity computational model. Results for surface roughness, and porosity from this analysis tool will help provide highly accurate part scale property predictions that consider the inherent variation in melt pool geometry in AM processes.

\section*{Acknowledgements}
\label{sec:sample1}
 The authors would like to acknowledge the support of NSF Grant CMMI-1934367. 

\appendix
\section{Kernel Density Estimation \citep{Davis2011}}
Kernel Density Estimation (KDE), a general and powerful way of estimating the probability density function of a random variable, is used to find the distribution of experimental data, which will be used to calibrate the stochastic simulation process parameters. The density function can be estimated through the first derivative of the distribution function. One of the simplest and most effective methods for estimating distribution functions is the so-called Empirical Distribution Function (EDF). That is, the estimate of $F_n\left(t\right)$ is the probability of all samples less than $t$:

 \begin{equation}
F_n\left(t\right)=\frac{The\ number\ of\ elements\ in\ the\ sample\ \le t}{n}=\frac{1}{n}\sum_{i=1}^{n}\mathbf{1}_{wi\le t} \label{eq22}
\end{equation}

\noindent where $\mathbf{1}_{wi\le t}$ is the indicator function. The indicator function of a subset A of a set $W$ is defined as:

\begin{equation}
\mathbf{1}_A\left(w\right)=\begin{cases}                                   

                   1\ \ \ if\ w\in\ A\\            

                   0\ \ \ if\ w\notin\ A\\             

                   \end{cases} \label{eq23}
\end{equation}

EDF is not differentiable and not smooth enough to compute the density function by the first derivative with respect to EDF. Thus, central difference can be used to find the density function:

\begin{equation}
f(w)=\lim \limits_{h\rightarrow 0}{\frac{F\left(w+h\right)-F(w-h)}{2h}} \label{eq24}
\end{equation}

Replace the distribution function with the empirical distribution function in Eq.\ref{eq22}, the numerator of Eq.\ref{eq24} is the number of points falling in the interval $[w-h,\ w+h]$, which can be written as:

\begin{equation}
f\left(w\right)=\frac{1}{2Nh}\sum_{i=1}^{N}{\mathbf{1}(w-h\le w_i\le w-h)}=\frac{1}{Nh}\sum_{i=1}^{N}{\frac{1}{2}\ast\mathbf{1}(\frac{\left|x-x_i\right|}{h}\le1)} \label{eq25}
\end{equation}

\noindent where $N$ is the number of sample points and h is the bandwidth. If so-called kernel function $K\left(t\right)=\frac{1}{2}\ast{\mathbf{1}}(t\le1)$ is used, Eq. \ref{eq25} can be further written as:

\begin{equation}
f\left(w\right)=\frac{1}{Nh}\sum_{i=1}^{N}{K(\frac{w-w_i}{h})} \label{eq26}
\end{equation}

Eq.\ref{eq26} gives the expression of KDE, which is also the estimation of probability density function. The integration of Eq.\ref{eq26} is:

\begin{equation}
\int{f\left(w\right)dw=\frac{1}{Nh}\sum_{i=1}^{N}\int{K(\frac{w-w_i}{h})}dw}=\frac{1}{N}\sum_{i=1}^{N}\int{K\left(t\right)dt=}\int K\left(t\right)dt
\end{equation}

Thus, as long as the integration of K is equal to 1, the integration of the estimated density function can be guaranteed as 1. The standard normal distribution can be used as kernel function, whose expression is:

\begin{equation}
K\left(t\right)=\frac{1}{\sqrt2\pi}e^{-\frac{t^2}{2}}
\end{equation}

Notice that the choice of h (also called bandwidth) in Eq.\ref{eq26} influence the goodness of KDE model. Here, Silverman’s rule of thumb algorithm is used as bandwidth selector due to its universality and effectiveness:

\begin{equation}
h=0.9\ast\ min\ {(\hat{\sigma},IQR/1.35)N}^{-\frac{1}{5}}
\end{equation}

\noindent where $\hat{\sigma}$ is the standard deviation of samples and $IQR$ is interquartile range (the difference between 75th and 25th percentiles).

\section{Kullback-Leibler Divergence \citep{10.1214/aoms/1177729694}}

Since both experimental and simulation results in section 4.3 are distributions $(f_{We}({{x}}), f_{De}\left({x}\right),f_{WPGD}({x}), f_{DPGD}({x}))$, Kullback-Leibler Divergence (KLD) can be used here to express difference between two continuous probability density distributions. KLD has its origins in the entropy of information theory, typically denoted as $H$. The definition of entropy for a probability distribution is:

\begin{equation}
H=-\int_{x}{p(x)\log(p\left(x\right))}dx
\end{equation}

\noindent where $p(x)$ is the probability density function of any random $x$. With the help of entropy, information can be quantified, and the loss of information can be measured when the observed distribution is substituted with parameterized approximation. Similarly, rather than just having probability distribution $p(x)$, KLD adds in the approximating distribution $q\left(x\right)$ and takes logarithm operation:

\begin{equation}
D_{KL}(p\left(x\right),q\left(x\right))=-\int_{x}{p(x)\log(p\left(x\right)-q\left(x\right)) }dx
\end{equation}

Essentially, KLD is the expectation of the log difference between the original (experimental) distribution with the approximating (simulated) distribution. A more common way to see KL divergence written is as follows:

\begin{equation}
D_{KL}(p\left(x\right),q\left(x\right))=\int_{x}{p(x)\log(\frac{q(x)}{p(x)})}dx
\end{equation}

%\section{}

%\nocite{*}
 \bibliographystyle{elsarticle-num} 
 \bibliography{main}





% \end{thebibliography}
\end{document}